# CO on a $Rh/Fe_3O_4$ single-atom catalyst: high-resolution infrared spectroscopy and near-ambient-pressure scanning tunnelling microscopy

Nail El Hocine Barama,[a] Chunlei Wang,[a] Panukorn Sombut,[a] David Rath,[a] Adam Lagin,[a] Martin Ormoš,[b] Lena Puntscher,[a] Faith J. Lewis,[a] Zdeněk Jakub,[a,c] Florian Kraushofer,[a] Moritz Eder,[a] Matthias Meier,[d] Michael Schmid,[a] Ulrike Diebold,[a] Cesare Franchini,[d,e] Peter Matvija,[b] Jiří Pavelec*[a] and Gareth S. Parkinson[a]

Infrared reflection absorption spectroscopy (IRAS) offers a powerful route to bridging the materials and pressure gaps between surface science and powder catalysis. Using a newly developed IRAS setup optimised for dielectric single crystals, we investigate CO adsorption on the model single-atom catalyst $Rh/Fe_3O_4(001)$. IRAS resolves three species: monocarbonyls at isolated, twofold-coordinated Rh adatoms, monocarbonyls at fivefold-coordinated Rh atoms embedded in the surface, and gem-dicarbonyls at isolated, twofold-coordinated Rh adatoms. Under ultra-high vacuum (UHV) conditions, RhCO monocarbonyl species at adatom sites dominate. $Rh(CO)_2$ gem-dicarbonyl formation is kinetically hindered and occurs predominantly through CO-induced dissociation of Rh dimers rather than sequential adsorption of two CO molecules at an isolated, twofold Rh adatom. The sequential-adsorption pathway to $Rh(CO)_2$ becomes accessible at millibar CO pressures as evidenced by near-ambient-pressure scanning tunnelling microscopy (NAP-STM). These findings link the UHV behaviour to that expected under realistic reaction conditions. Assignments of the vibrational frequencies to individual species rely on isotopic labelling, thermal treatments, and a review of previous SPM, XPS, and TPD data, supported by density functional theory (DFT)-based calculations. While theory provides qualitative insight, such as the instability of dicarbonyls on fivefold-coordinated Rh atoms, it does not yet reproduce experimental frequencies quantitatively and is sensitive to the computational parameters, highlighting the need for robust experimental benchmarks. The spectroscopic fingerprints established here provide a reliable foundation for identifying Rh coordination environments in oxide-supported single-atom catalysts.

## 1. Introduction

Infrared (IR) spectroscopy is one of the most powerful techniques in heterogeneous catalysis and is the most widely used among optical spectroscopies in this field,[1] as it can be used to monitor the evolution of surface-bound intermediates under operando conditions.[2–5] Furthermore, different implementations of this technique enable its application to diverse types of samples, ranging from powders, investigated by diffuse reflectance infrared Fourier transform spectroscopy (DRIFTS), to flat single crystals, studied by infrared reflection absorption spectroscopy (IRAS or IRRAS). As a result, IR studies on powder-supported catalysts have provided invaluable insight into reaction mechanisms. Moreover, CO titration is frequently used to study the properties of active sites because the CO-stretch frequency is sensitive to the local coordination environment.[6–14]

Oxide-supported Rh compounds are key materials in heterogeneous catalysis, for example, in automotive emission control and various selective hydrogenation and oxidation reactions.[15–19] Accordingly, IR absorption studies of CO adsorption on Rh-supported powder catalysts have attracted attention for decades. The earliest report dates back to 1957, when Yang and Garland investigated the sintering of dispersed Rh and assigned the resulting CO vibrational doublet to a single Rh adsorption site.[20] Since then, numerous studies have employed IR spectroscopy to investigate Rh-based powder catalysts, revealing both the complexity of CO adsorption and the diversity of the resulting surface species.[18,21–28] However, the structural complexity and heterogeneity of such catalysts make it difficult to unambiguously link a given vibrational feature to a specific atomic site. The coexistence of multiple adsorption geometries and the dynamic restructuring of active sites under reaction conditions further complicate interpretation, as highlighted in a recent review on IR characterisation of isolated atoms and nanoparticles.[29]

The surface-science approach enables experiments on well-defined catalytic model systems under tightly controlled UHV conditions. This includes single-atom catalysts (SACs) supported on single-crystal oxides, which provide a powerful platform for disentangling these complexities. Such model systems allow a direct correlation between atomic structure, electronic configuration, and vibrational signatures, whose frequencies are also, in principle, straightforward to calculate using DFT. Nonetheless, applying IRAS to metal oxide single crystals, the predominant support materials in catalysis, is challenging. The optimal incidence angles required to enhance adsorbate signals often coincide with regions of low infrared reflectivity, thereby drastically reducing signal intensity.[30] As a result, IRAS studies on metal oxides remain scarce. A common workaround is to grow thin oxide films on metal substrates to benefit from the high reflectivity of metals;[31–34] however, this strategy limits the accessible facets and may introduce structural or electronic deviations from those of bulk single crystals or conventional oxide powders. Overcoming these challenges for metal oxide single crystals is particularly important, as IRAS offers the unique combination of surface sensitivity and compatibility with elevated pressures, making it well suited to bridge the pressure gap between model catalysts and practical materials.[35,36]

While UHV-based surface-science studies provide insights at the atomic scale, they often cannot access adsorption pathways that are kinetically hindered at low pressure. In particular, adsorption steps that require short-lived intermediate geometries may be strongly suppressed under UHV conditions even when thermodynamically favourable. To assess how the CO adsorption mechanism evolves with pressure, and to test whether additional pathways become accessible under catalytic conditions, it is essential to complement UHV spectroscopy on well-investigated model systems with techniques capable of operating at elevated gas pressures. Adatoms of catalytic metals on the $Fe_3O_4(001)$ surface represent a versatile model system. $Fe_3O_4(001)$ exhibits a well-investigated subsurface cation-vacancy (SCV) reconstruction,[37] which exposes a periodic array of surface oxygen atoms that can stabilise isolated metal adatoms. This makes $Fe_3O_4(001)$ an ideal platform for SAC studies, and a wide range of metals, including Rh, have been systematically explored on this surface.[38–41] In the case of Rh, previous surface-science work has established how preparation conditions such as adsorption temperature, annealing, and coverage determine whether Rh occurs as isolated adatoms, substitutional species, or small clusters.[39,42–45] The interaction of CO with these Rh species has also been extensively characterised using experiments and DFT calculations, providing a detailed picture of the resulting adsorption geometries.[42,46,47] As a result, Rh/$Fe_3O_4(001)$ is one of the best-characterised model systems for oxide-supported Rh catalysts, offering atomic-level structural control that is difficult to achieve on powders. Investigating this system by IRAS is thus of particular interest, as it offers a direct link between fundamental surface-science insights and the behaviour of applied, high-surface-area catalysts.

In this work, we present a high-resolution IRAS investigation of CO adsorption on the model Rh/$Fe_3O_4(001)$ single-atom catalyst. By combining isotope-substitution experiments ($^{12}CO$, $^{13}CO$, and mixed $^{12}CO/^{13}CO$) with precise control over Rh coverage and surface preparation, we assign each observed vibrational feature unambiguously to a specific Rh coordination environment. DFT calculations capture the qualitative trends, but do not reproduce the vibrational frequencies quantitatively. Our IRAS study, therefore, provides a set of reference spectra for oxide-supported single-atom catalysts, offering a firm foundation for interpreting CO vibrational signatures in practical Rh catalysts and for evaluating theoretical predictions. To probe the pressure dependence of the CO adsorption mechanism, we further complemented the UHV-IRAS measurements with NAP-STM, which allowed us to determine whether the sequential-adsorption pathway to $Rh(CO)_2$, kinetically inaccessible in UHV, becomes accessible under mbar CO exposures.

## 2. Methods

IRAS measurements were performed using a newly developed setup designed to optimise to maximise the signal-to-noise ratio for metal oxide surfaces[48]. All spectra were acquired with a resolution of 4 $cm^{-1}$ and averaged over 4000 scans, corresponding to an acquisition time of approximately 20 min per spectrum. The measurements were conducted with p-polarised light at non-grazing incidence angles between 55° and 74°, a range optimised for the $Fe_3O_4(001)$ surface to enhance peak intensity. The spectrometer used in the IRAS setup is a Bruker VERTEX 80v. The entire system is housed within a custom-built UHV chamber designed for chemical reactivity studies, operating at a base pressure below $1 \times 10^{-10}$ mbar.[49] The chamber is equipped with a triple-pocket electron-beam evaporator for metal deposition and a temperature-stabilised quartz-crystal microbalance (QCM) used to monitor and calibrate the Rh deposition rate. For $Fe_3O_4(001)$, one monolayer (1 ML) is defined as one Rh atom per $(\sqrt{2} \times \sqrt{2})R45°$ unit cell, corresponding to $1.42 \times 10^{14}$ atoms per $cm^2$. Carbon monoxide was dosed using a home-built molecular beam source that provides uniform exposure over a well-defined ≈3.5 mm diameter area on the sample surface.[49,50] The UHV chamber is equipped with several complementary surface-science techniques, including X-ray photoelectron spectroscopy (XPS), ultraviolet photoelectron spectroscopy (UPS), temperature-programmed desorption (TPD), low-energy electron diffraction (LEED), and low-energy ion scattering (LEIS). In the present work, XPS was employed to verify surface cleanliness and confirm Rh coverages. XPS spectra were recorded using a monochromatized Al/Ag twin-anode X-ray source (SPECS XR50 M, FOCUS 500) and a hemispherical analyser (SPECS Phoibos 150) with Al Kα radiation.
UHV scanning tunnelling microscopy (STM) measurements were performed on a separate $Fe_3O_4$(001) sample in an independent UHV system consisting of a preparation chamber ($p$ < 1 × $10^{-10}$ mbar) and an analysis chamber ($p$ < 7 × $10^{-11}$ mbar). The STM (Omicron μ-STM) was operated at room temperature in constant-current mode using electrochemically etched tungsten tips. Further details of this setup are provided in previous papers.[38,51,52]

Non-contact atomic force microscopy (nc-AFM) measurements were carried out on an additional $Fe_3O_4$(001) sample in a dedicated UHV chamber ($p$ < 1 × $10^{-10}$ mbar) using a qPlus sensor[53] ($f_0$ = 31.8 kHz, $k$ = 1800 N/m, $Q$ ≈ 10 000) with an electrochemically etched tungsten tip. The CO-tip functionalization was spontaneously formed during STM scanning or mild bias pulsing above Rh–CO species, similarly to previous reports.[40]

Near-ambient-pressure measurements were performed at the Nanomaterials Group at Charles University in Prague. The system consists of a UHV preparation chamber equipped with XPS, an electron-beam evaporator, an electron beam heating stage, and an oxygen supply, connected via a transfer line to a second chamber housing an Aarhus NAP-STM/AFM microscope. The base pressure was 1 × $10^{-9}$ mbar in the preparation chamber and 1 × $10^{-10}$ mbar in the STM/AFM chamber. CO (Linde, 99.997%) was introduced into the STM/AFM enclosure via a leak valve equipped with a cryogenic ($LN_2$-cooled) filter. Near-ambient pressures in the range of 1–10 mbar were applied at room temperature and monitored using a calibrated Pirani gauge. STM images were acquired in constant-current mode under stable CO pressure, with typical tunnelling conditions of +0.7–1.2 V sample bias and 20–50 pA. After the NAP-STM measurements, the high-pressure cell was pumped back to UHV, and the sample was imaged again to confirm that no pressure- or tip-induced morphological changes had occurred.
All samples used in this study were natural $Fe_3O_4$(001) single crystals obtained from SurfaceNet GmbH. Surface preparation consisted of repeated cycles of $Ne^+$ (or $Ar^+$, depending on the chamber) ion sputtering at 300 K (1 keV, 15 min) followed by annealing under UHV at 900 K for 15 min. Every second annealing cycle was performed in an oxygen atmosphere ($p(O_2)$ = 3 × $10^{-7}$ mbar, 20 min) to prevent reducing the sample.

VASP was used for all DFT calculations with the PAW method and a 550 eV plane-wave cutoff.[54–56] Electronic exchange-correlation was treated using the GGA-PBE functional with D3 dispersion corrections (Becke-Johnson damping), and strong correlation of Fe 3d states was described with $U_{eff}$ = 3.61 eV.[57–60] Electronic self-consistency and ionic-relaxation were converged to $10^{-6}$ eV and 0.02 eV/Å, respectively. The $Fe_3O_4$(001) surface was modelled by an asymmetric 13-layer slab (7 octahedral and 6 tetrahedral Fe layers) in a $(2\sqrt{2}\times2\sqrt{2})R45°$ supercell, with the top 4 layers relaxed, the bottom 9 fixed, Γ-point sampling only, and 14 Å of vacuum.[46] CO vibrational frequencies were computed in the harmonic approximation using finite differences. Motivated by the use of DFT+U to mitigate self-interaction errors in localised d states, we also examined the effect of applying a Hubbard $U$ to the Rh 4d states, using the same $U_{eff}$ value as for the Fe 3d states, on the calculated CO stretching frequencies. The computed CO stretching vibrational frequency for adsorbed CO on the Rh/$Fe_3O_4$(001) was scaled by the factor $\lambda = v^{exp}_{CO_{gas}}/v^{cal}_{CO_{gas}}$ to compensate for the systematic DFT errors, where $v^{exp}_{CO_{gas}}$ is the frequency of an isolated CO molecule in the gas phase.[61]

## 3. Summary of results from our previous studies

The adsorption of Rh on $Fe_3O_4$(001) in the low-coverage regime has been comprehensively characterised in our previous studies.[42,46,47] Additional work by Sharp *et al.* further refined this picture by mapping the distribution of Rh species over a wide range of coverages and annealing temperatures.[39] In the present study, we focus on low Rh coverages (< 0.5 ML) deposited at room temperature, where isolated Rh adatoms dominate. These adatoms are twofold-coordinated ($Rh_{2fold}$) to surface oxygen atoms within the SCV reconstruction and appear in STM images as bright protrusions centred between the rows of octahedrally coordinated Fe atoms ($Fe_{oct}$).[42] The adsorption energy of $Rh_{2fold}$ was calculated to be $E_{ads}$ = −4.42 eV.[46] Occasionally, $Rh_{2fold}$ adatoms pair to form $Rh_2$ dimers, which appear as elongated protrusions in STM. At room temperature, the dimers between two equivalent configurations by moving parallel to the $Fe_{oct}$.[42,46] DFT calculations indicate that these dimers are thermodynamically preferred over two isolated adatoms by approximately 0.19 eV, but kinetic barriers prevent pairing of all $Rh_{2fold}$ adatoms at room temperature. Dimers account for approximately 12 % of the total Rh population at an initial coverage of 0.2 ML. In addition to adatoms and dimers, some Rh atoms substitute surface/subsurface $Fe_{oct}$ sites within the spinel structure to form fivefold ($Rh_{5fold}$) or sixfold ($Rh_{6fold}$) coordinated Rh sites.[42] These substitutional Rh atoms are stabilised relative to the $Rh_{2fold}$ adatom by 0.67 eV for the fivefold configuration and by 1.10 eV for the sixfold subsurface configuration. Only the surface $Rh_{5fold}$ species are relevant to our CO adsorption studies here. Although these substitutional Rh atoms are already present in small quantities after room-temperature deposition, annealing above ≈400 K drives a substantial fraction of $Rh_{2fold}$ adatoms into the more highly coordinated $Rh_{5fold}$ sites.

Figure 1 summarises the three principal CO adsorption configurations on Rh sites observed after CO exposure on a low-coverage Rh/$Fe_3O_4$(001) surface.[42,46] Figure 1a presents an STM image (top) and a nc-AFM image (bottom) of the same surface region, both recorded at 78 K using a CO-terminated tip after CO exposure. The most abundant species are $Rh_{2fold}CO$ monocarbonyls (highlighted by cyan arrows).[46] Although CO adsorption modifies the local coordination of the Rh adatom, we retain the $Rh_{2fold}$ notation throughout for clarity, referring to the coordination of the Rh site prior to CO adsorption. Upon CO adsorption, the $Rh_{2fold}$ adatom sinks slightly into the surface and forms an additional bond with one of the two equivalent subsurface oxygen atoms, forming a pseudo-square-planar geometry tilted away from the surface normal. The weak bond to the subsurface oxygen enables rapid switching between two equivalent configurations along the [110] direction. The calculated switching barrier is approximately 0.1 eV.[46] As a result, at room temperature, the species appear as faint double protrusions in nc-AFM and as elongated features in STM. The corresponding structural model and calculated adsorption energy are shown in Figure 1b.

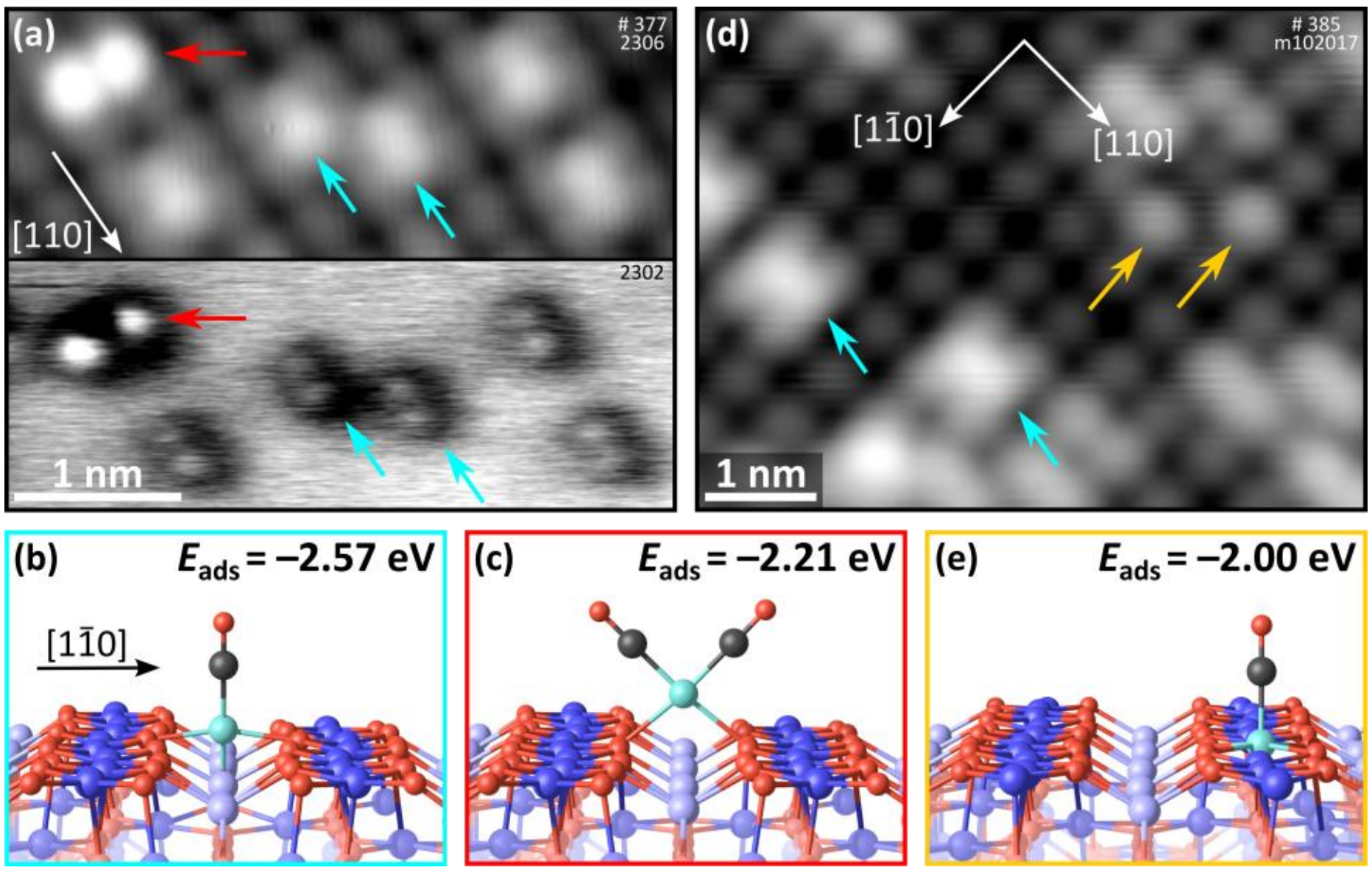

**Figure 1. CO adsorption configurations on Rh/$Fe_3O_4$(001).** (a) STM (top) and nc-AFM (bottom) images acquired at 78 K after CO exposure on low-coverage Rh/$Fe_3O_4$(001). Cyan arrows mark monocarbonyls on $Rh_{2fold}$ sites, and red arrows indicate gem-dicarbonyls on $Rh_{2fold}$ sites. (b) DFT model of the pseudo-square-planar geometry adopted by the $Rh_{2fold}$ monocarbonyl. The CO is not vertical but tilted towards the viewer by ≈20°. (c) DFT model of the gem-dicarbonyl on a $Rh_{2fold}$ site. (d) STM image extracted from a movie recorded during CO exposure on 0.2 ML Rh/$Fe_3O_4$(001). Here, 1 monolayer (ML) corresponds to $1.42 \times 10^{14}$ Rh atoms per $cm^2$, i.e. one Rh atom per ($\sqrt{2} \times \sqrt{2}$)R45° surface unit cell of $Fe_3O_4$(001). Yellow arrows mark CO adsorbed on $Rh_{5fold}$ atoms substituting $Fe_{oct}$ sites. (e) DFT model of CO adsorbed on a $Rh_{5fold}$ site. Adsorption energies are based on PBE-D3 with $U_{Rh} = 0$.

In addition, the bright double-lobed protrusions oriented perpendicular to the $Fe_{oct}$ rows (red arrows in Figure 1a) are assigned to $Rh_{2fold}(CO)_2$ gem-dicarbonyl species. As demonstrated by Wang *et al.*,[46] these dicarbonyls form almost exclusively through CO-induced dissociation of the minority $Rh_2$ dimers present on the surface under UHV conditions. This process was captured in sequential STM images, and DFT calculations revealed that, upon CO exposure, the $Rh_2$ dimers pass through an unstable intermediate $Rh_2(CO)_3$ configuration, which subsequently dissociates to produce one monocarbonyl $Rh_{2fold}CO$ and one gem-dicarbonyl $Rh_{2fold}(CO)_2$. Although the $Rh_{2fold}(CO)_2$ configuration is thermodynamically stable (average adsorption energy $E_{ads}$ = −2.37 eV per CO with the PBE-D3 functional used in the current work; no $U$ on Rh), sequential adsorption of two CO molecules on an isolated $Rh_{2fold}$ adatom was never observed in UHV. Our previous combined STM/DFT analysis suggested that this pathway requires the $Rh_{2fold}CO$ to adopt a transient, sterically constrained geometry with an extremely low configurational probability. Since the $Rh_{2fold}CO$ assumes this configuration only during a tiny fraction of the time (<$10^{-8}$), adsorption of another CO during these short periods is extremely unlikely at UHV gas impingement rates.[46] This kinetic picture predicts that sufficiently high CO pressures should enable the sequential-adsorption mechanism by dramatically increasing the rate at which CO molecules encounter $Rh_{2fold}CO$ species in the required transient geometry. As shown in section 5, NAP-STM measurements presented in the current work indeed reveal that the sequential CO adsorption pathway becomes accessible under mbar CO pressures, providing direct experimental support for this kinetic interpretation. The structural model of $Rh_{2fold}(CO)_2$ and the calculated adsorption energy per CO molecule are shown in Figure 1c.

CO can also adsorb on the substitutional $Rh_{5fold}$ sites. In this case, DFT calculations indicate that gem-dicarbonyl formation is energetically unfavourable.[46] Only a single CO molecule can bind to a $Rh_{5fold}$ atom, completing an octahedral coordination environment (Figure 1e). CO adsorption at these sites is significantly weaker ($E_{ads}$ = −2.19 eV with PBE-D3, $U_{Rh}$ = 0) than at the $Rh_{2fold}$ sites ($E_{ads}$ = −2.52 eV).[46] These calculations are in excellent agreement with CO TPD experiments, where monocarbonyls on $Rh_{5fold}$ lead to a desorption peak at around 410 K, notably lower than the desorption peak stemming from the $Rh_{2fold}$ at 530 K.[42] The STM image in Figure 1d is extracted from a movie recorded before and during CO exposure (included as Supporting Information, Movie S1). The movie illustrates how $Rh_{2fold}$ sites transform into elongated features upon CO adsorption (highlighted again by cyan arrows).[46] In contrast, the $Rh_{5fold}$ sites exhibit only a subtle change in contrast upon CO adsorption, and the CO does not appear elongated (no flipping between two configurations; yellow arrows in Figure 1d). Owing to their low concentration under the present deposition conditions (temperature and coverage), these species are observed only rarely in STM. Nevertheless, they are expected to give distinct signatures in the infrared spectra, as IRAS is very sensitive to dipole moments normal to the surface. In principle, low-temperature infrared spectroscopy could also be used to probe dimer carbonyl intermediates, provided the dimers are stabilized under the chosen conditions. In the present work, we focus on the three adsorption configurations observed at room temperature. Together, these three configurations—monocarbonyls on $Rh_{2fold}$, gem-dicarbonyls on $Rh_{2fold}$, and monocarbonyls on $Rh_{5fold}$—constitute the structural basis for interpreting the IRAS spectra discussed in section 4. Their distinct adsorption geometries and adsorption energies are expected to produce characteristic shifts in the CO stretching frequencies that allow each configuration to be distinguished spectroscopically, and they represent the three relevant Rh–CO states from which the kinetic behaviour under different CO pressures emerges. Section 5 provides a direct NAP-STM test of this kinetic picture.

## 4. IRAS of CO on Rh/$Fe_3O_4$(001)

## 4.1. Three different CO species

To enable a direct comparison with the STM and DFT results summarised in Figure 1, IRAS measurements were performed on a Rh-decorated $Fe_3O_4$(001) sample with Rh coverages that had been previously characterised in detail.[42,46] Figure 2 shows p-polarised IRAS spectra acquired after exposure to 3.4 langmuirs (L; 1 L = 1.33 x $10^{-6}$ mbar.s) of $^{13}$CO at room temperature for two different Rh coverages, 0.2 ML (black spectrum) and 0.4 ML (blue spectrum). In all experiments, the reference spectrum $R_0$ was recorded after Rh deposition but prior to CO dosing at 300 K, ensuring that only CO-induced changes appear in the spectra. For 0.2 ML Rh, three distinct absorption bands are observed in the CO-stretching region at 1979, 2037, and 2059 cm$^{-1}$. All three features are also present at 0.4 ML Rh and increase in intensity with increasing Rh coverage, indicating that each originates from CO adsorbed on Rh-related sites. The positions of the three bands remain nearly unchanged in the 0.4 ML spectrum, which suggests that dipole–dipole interactions between neighbouring CO molecules are negligible. This is consistent with the previous STM studies,[42,46] which show that the nearest-neighbour distance between $Rh_{2fold}$ adatoms on $Fe_3O_4$(001) is fixed by the surface reconstruction at 8.4 Å.

Based on the CO stretch frequencies calculated by DFT (Table 1), we assign the low-frequency peak at 1979 cm$^{-1}$ to CO adsorbed on the $Rh_{2fold}$ sites (Fig. 1b). The relatively high intensity of this band is consistent with STM results, which show that $Rh_{2fold}CO$ species are by far the most abundant configuration containing CO under these conditions.[46] We note, however, that IRAS peak intensities are not solely determined by the population of a given species, but also depend on the magnitude and orientation of the dynamic dipole moment relative to the surface normal, and on the detailed bonding geometry.

**Table 1.** Experimental and calculated C–O stretching frequencies, together with calculated CO adsorption energies for the different adsorption configurations. The calculated frequencies are obtained by DFT with and without Hubbard *U* correction applied to the Rh 4d states.

| Configuration | $E_{ads}$, CO (eV) | $\tilde{\nu}_{DFT}$, $^{13}$CO (cm$^{-1}$) | | $\tilde{\nu}_{exp}$, $^{13}$CO (cm$^{-1}$) | $\tilde{\nu}_{DFT}$, $^{12}$CO (cm$^{-1}$) | | $\tilde{\nu}_{exp}$, $^{12}$CO (cm$^{-1}$) |
|---|---|---|---|---|---|---|---|
| | | $U_{eff, Rh} = 0$ | $U_{eff, Rh} = 3.61$ eV | | $U_{eff, Rh} = 0$ | $U_{eff, Rh} = 3.61$ eV | |
| **$Rh_{2fold}CO$** | −2.52 | 1971 | 1978 | 1979 | 2021 | 2029 | 2026 |
| **$Rh_{2fold}(CO)_2$** | −2.37 | 2022 (sym) | 2023 (sym) | 2037 | 2076 (sym) | 2076 (sym) | 2085 |
| | | 1961 (asym) | 1974 (asym) | | 2012 (asym) | 2025 (asym) | |
| **$Rh_{5fold}CO$** | −2.19 | 2025 | 2047 | 2059 | 2078 | 2100 | 2106 |

The two higher-frequency bands at 2037 cm$^{-1}$ and 2059 cm$^{-1}$ appear with much lower intensity than the dominant 1979 cm$^{-1}$ band in both spectra. As mentioned previously, only small populations of $Rh_2$ dimers (the precursors of gem-dicarbonyls) and $Rh_{5fold}$ sites are present at this coverage. This low abundance is reflected in the correspondingly weak intensity of these two features. Based on DFT, the assignment of these two peaks is uncertain since the calculation without $U$ on the Rh 4d states yields almost identical frequencies for the symmetric mode of the gem-dicarbonyl and the CO at the $Rh_{5fold}$ embedded in the surface. In the following, we will demonstrate that the assignment of the experimental IR bands in Figure 2 and Table 1 is correct, although the agreement between DFT and experiment is worse than for the $Rh_{2fold}CO$.

In principle, one could also consider whether one of the two high-frequency IRAS peaks originates from CO adsorbed on small Rh clusters. However, at a coverage <0.5 ML Rh, the amount of Rh clusters is negligible.[39] Moreover, because clusters exhibit a wide distribution of sizes and geometries, their vibrational signatures are typically broad rather than sharp, unlike the well-defined peaks observed in Fig. 2. Finally, CO adsorbed on small Rh clusters normally exhibits lower stretching frequencies (≈2040 cm$^{-1}$ for $^{12}$CO).[62]

Table 1 shows that the CO stretch frequencies are directly correlated to the adsorption energies. A strong Rh-CO interaction is associated with back-donation from Rh 4d into the CO 2π* orbital, weakening the C–O bond and shifting its stretching frequency to lower values.[63–66]

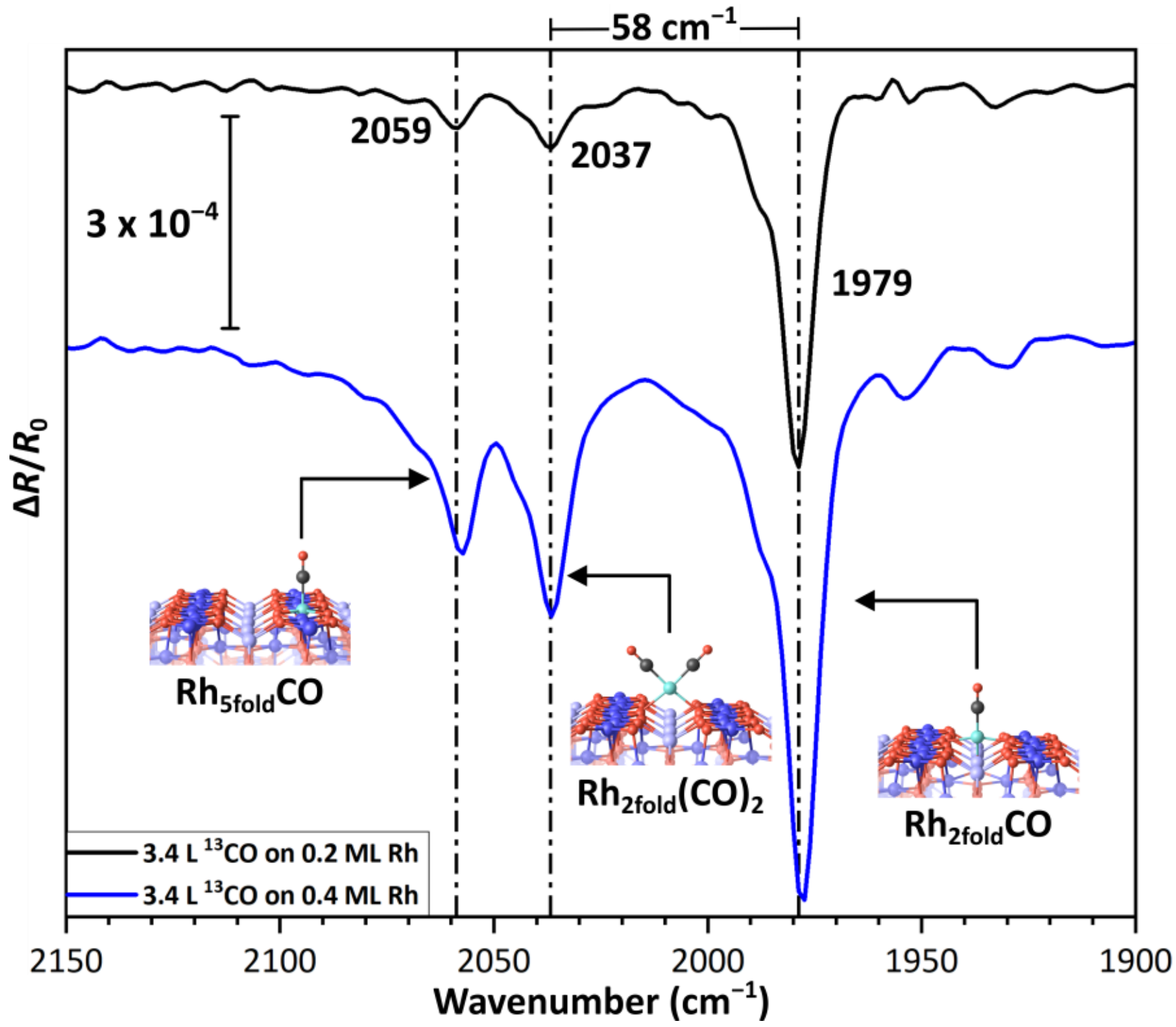

**Figure 2. High-resolution IRAS spectra of $^{13}$CO adsorbed on Rh/$Fe_3O_4$(001).** p-polarised IRAS spectra of Rh-decorated $Fe_3O_4$(001) with Rh coverages of 0.2 ML (black) and 0.4 ML (blue) after 3.4 L $^{13}$CO exposure. $^{13}$CO was introduced to the surface at room temperature, and the IRAS spectra were collected under identical conditions at $T$ = 300 K. Three distinct vibrational bands are observed at 1979, 2037, and 2059 cm$^{-1}$; their intensities increase with Rh coverage. These features are assigned to CO adsorbed on twofold-coordinated $Rh_{2fold}$ sites, the symmetric stretch of gem-dicarbonyl $Rh_{2fold}(CO)_2$ species, and CO adsorbed on $Rh_{5fold}$ sites,

respectively. Spectra were acquired with 4 $cm^{-1}$ resolution and averaged over 4000 scans (≈20 min acquisition time per spectrum).

## 4.2. Annealed $Rh/Fe_3O_4(001)$: Increase of $Rh_{5fold}$ sites

As shown in previous STM and XPS studies[39,42], $Rh_{2fold}$ adatoms are stable only below 400 K. Upon heating above this temperature, these species begin to incorporate into the surface lattice, forming fivefold and sixfold Rh sites. This structural transformation should be detectable by IRAS, as CO molecules adsorbed on twofold and fivefold Rh sites exhibit distinct C–O stretching frequencies, whereas $Rh_{6fold}$ is below the surface and therefore not capable of CO adsorption. Figure 3 presents p-polarised IRAS spectra of $Rh/Fe_3O_4(001)$ surfaces that all had the same initial Rh coverage of 0.2 ML but underwent different thermal treatments prior to 3.4 L $^{13}CO$ dosing, leading to a different distribution of Rh species. The black spectrum corresponds to the as-deposited sample, which was kept at room temperature, and is identical to the one shown in Figure 2.

For the orange spectrum, the sample was briefly flashed to ≈420 K after Rh deposition and then cooled back to 300 K. The spectrum shows a pronounced increase in the intensity of the high-frequency band at 2059 $cm^{-1}$, consistent with its assignment to monocarbonyls on $Rh_{5fold}$ sites that form upon thermal conversion of $Rh_{2fold}$ species. The accompanying decrease in the low-frequency peak further supports this interpretation. The shoulder of the peak at 1979 $cm^{-1}$ is attributed to CO adsorbed on $Rh_{2fold}$ adatoms located at defect sites, primarily antiphase domain boundaries (APDBs) and sites above additional subsurface $Fe_{oct}$ cations, as described in our previous studies[67,68]. Such defect sites can also stabilise Rh adatoms in a twofold-coordination with a slightly higher adsorption energy, resulting in a weaker Rh–CO interaction and consequently a blue shift of the C–O stretching frequency. The shoulder appears more distinct in the orange spectrum mainly because the 1979 $cm^{-1}$ peak undergoes a slight red shift after flashing to ≈420 K. A similar small shift is observed for the intermediate band at 2037 $cm^{-1}$. These small shifts likely reflect subtle changes in the Rh–CO interaction strength due to modest modifications in the electronic structure of the surface and subsurface layers. Such changes can accompany structural rearrangements in which some $Rh_{2fold}$ adatoms convert into fivefold surface species, while others incorporate into subsurface positions to form $Rh_{6fold}$, as seen previously for Rh and other transition metal adatoms [38,42]. The nearly unchanged intensity of the shoulder compared to the black spectrum indicates that regular twofold $Rh_{2fold}$ species are the first to convert into $Rh_{5fold}$, while those associated with defects remain stable after this brief heat treatment.

The red spectrum in Figure 3 was recorded at 300 K after annealing the surface for 5 minutes at 420 K. This longer temperature treatment further decreases the intensity of both the low-frequency $Rh_{2fold}CO$ band and its shoulder, while the high-frequency band assigned to CO on $Rh_{5fold}$ becomes more intense. This provides direct experimental support for the peak assignments, as it is consistent with earlier STM and XPS observations of the evolution of Rh species under comparable annealing conditions.[39,42] In contrast to the behaviour of the twofold band, the fivefold band shows a slight blue shift after annealing, likely reflecting a modest reduction in π-backdonation from $Rh_{5fold}$ to CO as they become more tightly integrated into the surface lattice. The resulting weaker Rh–CO bond leads to a slightly higher C–O stretching frequency, in agreement with the observed shift.

Although the shift in intensity from the low- to the high-frequency band provides clear qualitative evidence for their assignment to CO on twofold and fivefold Rh sites, respectively, it remains difficult to determine quantitatively how many $Rh_{2fold}$ atoms convert into $Rh_{5fold}$ upon annealing. This is because CO binds in different geometries on the two sites, leading to different orientations of the vibrational dipole moment relative to the surface, and, possibly, different magnitudes of the dipole moment. CO adsorbed on $Rh_{5fold}$ stands upright, so its dipole moment change is normal to the surface and couples strongly to the perpendicular component of p-polarised light. In contrast, CO bound to $Rh_{2fold}$ adopts a pseudo-square-planar geometry and is tilted, resulting in both perpendicular and in-plane components of the dipole moment. Our measurements were conducted with p-polarised light at non-grazing incidence angles below 74°. At these incidence angles, perpendicular dipoles produce strong negative features, whereas in-plane dipoles give much weaker positive signals (approximately one-fifteenth of the perpendicular intensity if the dipole is parallel to the incidence plane, and no signal in p polarisation for the case of a dipole moment normal to the incidence plane; see Figure S1). Consequently, the intensities of the two peaks are not a quantitative measure of the abundance of these species.

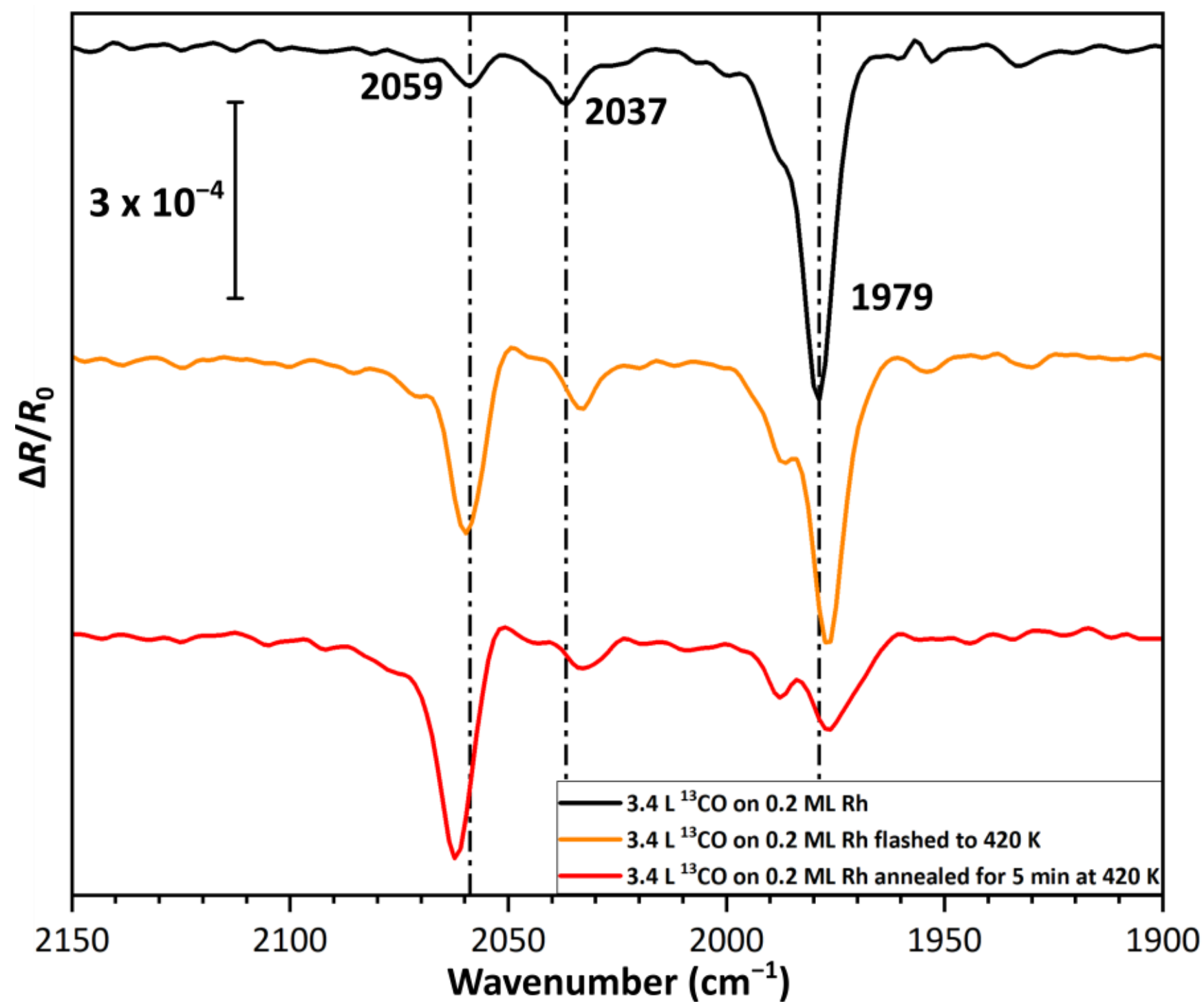

**Figure 3. Thermally induced conversion of twofold to fivefold Rh sites monitored by IRAS:** p-polarised IRAS spectra of Rh-decorated $Fe_3O_4(001)$ surfaces (initial Rh coverage 0.2 ML) subjected to different thermal treatments prior to 3.4 L $^{13}CO$ exposure. The black spectrum corresponds to the as-deposited sample kept at 300 K (same as Figure 2). The orange spectrum was obtained after a brief flash to 420 K followed by cooling to 300 K, while the red spectrum was recorded after annealing the surface for 5 min at 420 K. All spectra were measured at 300 K after CO exposure.

## 4.3. Isotopic confirmation of the gem-dicarbonyl band

To provide direct experimental confirmation of the assignment of the mid-frequency peak to gem-dicarbonyl $Rh_{2fold}(CO)_2$ species, isotopic substitution experiments were performed using $^{12}CO$, $^{13}CO$, and an equimolar $^{12}CO/^{13}CO$ mixture on identically prepared 0.4 ML $Rh/Fe_3O_4(001)$ surfaces. Figure 4 presents the corresponding p-polarised IRAS spectra. The top (blue) spectrum corresponds to $^{13}CO$, identical to that shown in Figure 2; the bottom (purple) to $^{12}CO$; and the middle (dark violet) to the $^{12}CO/^{13}CO$ mixture.
As expected, the pure $^{12}CO$ and $^{13}CO$ spectra each display three absorption bands of comparable relative intensity, with all $^{13}CO$ features shifted to lower frequencies by ≈47 cm$^{-1}$ relative to the other isotope. This isotopic shift is consistent with the change in the reduced mass of the C–O bond and is in excellent agreement with the harmonic-oscillator model. For monocarbonyl species adsorbed on Rh atoms, exposure to an equimolar mixture of $^{12}CO$ and $^{13}CO$ should yield two peaks of similar intensity separated by ≈47 cm$^{-1}$, corresponding to identical Rh sites bonded to either isotope. This behaviour is indeed observed in the mixed-isotope (dark violet) spectrum for the two peaks at 1979 and 2059 cm$^{-1}$ with $^{13}CO$ (2106 and 2026 cm$^{-1}$ with $^{12}CO$), confirming their assignment as monocarbonyl Rh–CO species. In contrast, gem-dicarbonyls behave differently: co-adsorption of $^{12}CO$ and $^{13}CO$ should produce a third, intermediate band between the pure-isotope peaks, arising from mixed-isotope $Rh_{2fold}(^{12}CO)(^{13}CO)$ species. The appearance of such an intermediate band is a well-established fingerprint of intramolecular vibrational coupling between the two CO ligands, as reported previously by Frank *et al.*[26]

In the present case, the mixed-isotope spectrum reveals a new feature at 2068 cm$^{-1}$, providing solid evidence that the mid-frequency band in the pure-isotope spectra originates from $Rh_{2fold}(CO)_2$ rather than from distinct monocarbonyl species. The measured peak separations between the symmetric $^{12}CO$ and mixed-isotope bands (≈17 cm$^{-1}$), and between the mixed-isotope and $^{13}CO$ bands (≈31 cm$^{-1}$), are fully consistent with the previously reported coupling strengths for gem-dicarbonyls on Rh atoms supported on NiAl(110).[26] Furthermore, the approximately two-fold higher intensity of the mixed-isotope 2068 cm$^{-1}$ band relative to the pure $^{13}CO$ dicarbonyl feature is fully consistent with statistical expectations: in an equimolar $^{12}CO/^{13}CO$ mixture, the probability of forming $Rh_{2fold}(^{13}CO)(^{12}CO)$ is twice that of forming $Rh_{2fold}(^{13}CO)_2$ or $Rh_{2fold}(^{12}CO)_2$.

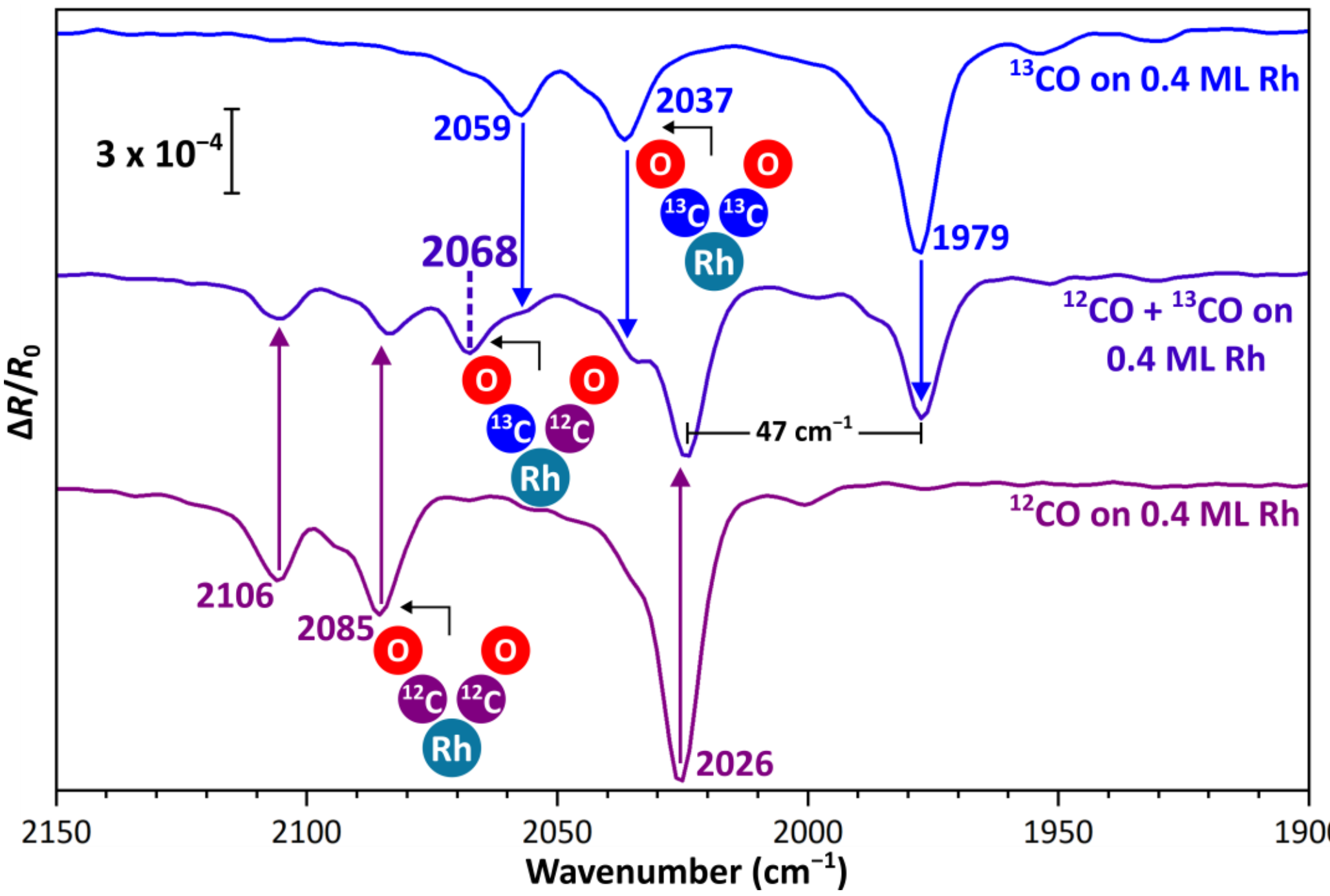

**Figure 4. Isotopic confirmation of the gem-dicarbonyl assignment.** p-polarized IRAS spectra recorded at 300 K after exposing 0.4 ML Rh/$Fe_3O_4$(001) to $^{13}$CO (blue), $^{12}$CO (purple), and an equimolar $^{12}$CO/$^{13}$CO mixture (dark violet). Each pure-isotope spectrum exhibits three bands corresponding to CO adsorbed on twofold and fivefold Rh sites, as well as the symmetric stretch of the $Rh_{2fold}(CO)_2$ gem-dicarbonyl. In the $^{12}$CO spectrum, all $^{13}$CO features are shifted to lower frequency by ≈47 cm$^{-1}$, consistent with the expected isotopic dependence of the C–O stretching mode. In the mixed-isotope experiment, an additional intermediate band appears between the pure-$^{12}$CO and pure-$^{13}$CO gem-dicarbonyl peaks, arising from vibrational coupling within mixed-isotope $Rh_{2fold}(^{13}CO)(^{12}CO)$ species at 2068 cm$^{-1}$, see sketch. The presence of this intermediate feature provides unambiguous evidence that the mid-frequency band at ≈2037 cm$^{-1}$ (for $^{13}$CO) originates from a gem-dicarbonyl species rather than a monocarbonyl. The relative intensities of the three gem-dicarbonyl peaks in the mixed-isotope spectrum are consistent with the expected 1:2:1 statistical distribution for an equimolar $^{12}$CO/$^{13}$CO mixture.

Gem-dicarbonyls exhibit two characteristic C–O stretching modes, a symmetric and an asymmetric stretch,[18,22,27,69,70] but we have only discussed the symmetric stretch so far. In powder catalysts, both bands appear as minima in relative reflectivity (i.e., upward absorbance peaks) because the crystallites are randomly oriented with respect to the incident IR light. On a single-crystal surface, the sign and magnitude of $\Delta R/R_0$ depend on dipole orientation, light polarisation, and incidence angle, as described by the Fresnel equations (Figure S1). Perpendicular dipole components give strong negative bands, whereas in-plane dipoles generate much weaker positive features. The symmetric stretch of the $Rh_{2fold}(CO)_2$ gem-dicarbonyl has a net dipole-moment change normal to the surface and therefore appears as the negative peak (2037 cm$^{-1}$ for $^{13}$CO). In contrast, the asymmetric stretch exhibits an in-plane dipole moment change and would appear as a very weak, positive band. In a recent IRAS study of Rh gem-dicarbonyls on $TiO_2$(110), Eder *et al.*[62] observed the asymmetric stretch as a weak positive peak ≈61 cm$^{-1}$ below the symmetric stretch peak, in excellent agreement with DFT predictions. Thus, in our spectra, the asymmetric stretch would be extremely weak compared with the symmetric stretch at 2037 cm$^{-1}$ and would fall below the noise level. Moreover, its expected frequency coincides almost exactly with the intense $Rh_{2fold}CO$ monocarbonyl band at 1979 cm$^{-1}$, which lies 58 cm$^{-1}$ below the symmetric stretch—very close to the offset reported by Eder *et* al.[62] and to values known from powder catalysts.[22,69,70]

The 0.4 ML spectrum in Figure 2 shows that the relative intensities of both the high-frequency band at 2059 cm$^{-1}$ and the intermediate band at 2037 cm$^{-1}$ increase more strongly with Rh coverage than that of the dominant $Rh_{2fold}CO$ band (See Table S1). This behaviour is fully consistent with our peak assignment. For the $Rh_{5fold}$ embedded in the surface, it has been shown that its occurrence markedly increases with an increase of the Rh coverage to 0.5 ML.[39] Likewise, $Rh_2$ dimers become more abundant at higher Rh coverages. Initially, deposited Rh atoms diffuse until they are trapped at isolated twofold sites; with increasing coverage, a larger fraction of these sites is already occupied, increasing the probability that an incoming Rh atom encounters and binds to an existing $Rh_{2fold}$ adatom. The resulting dimers are thermodynamically stable, being ≈0.19 eV more stable than isolated $Rh_{2fold}$ adatoms,[46] which leads to a higher concentration of $Rh_{2fold}(CO)_2$ gem-dicarbonyls upon CO exposure under UHV conditions. To determine whether gem-dicarbonyls remain a minority species also at elevated pressures, we carried out complementary near-ambient-pressure STM measurements, which are presented in the next section.

## 5. Near-ambient-pressure STM: sequential CO adsorption at elevated pressure

As described in section 3, in UHV experiments, dicarbonyls can form only via dissociation of $Rh_2$ dimers.[46] To test whether this restriction persists at higher pressures, we examined CO adsorption on $Rh/Fe_3O_4(001)$ using NAP-STM. After standard UHV preparation and a brief low-dose CO exposure (≈3 L), STM imaging confirmed that the surface contained primarily monocarbonyl $Rh_{2fold}CO$ species, as in all previous UHV studies. When a much higher CO dose was applied, 2 mbar for 3 min at room temperature, the surface changed dramatically. STM images recorded at 2 mbar CO [Fig. 5(a)] showed that essentially all $Rh_{2fold}CO$ species adopt a double-lobed geometry identical to the $Rh_{2fold}(CO)_2$ gem-dicarbonyls observed in UHV only after CO-induced dissociation of $Rh_2$ dimers [red arrows in Fig. 5(a)]. Under these near-ambient conditions, however, the gem-dicarbonyls appear in the absence of any monocarbonyl neighbours (which would result from the Rh-dimer mediated mechanism of dicarbonyl formation), demonstrating that the dicarbonyls now form directly from isolated $Rh_{2fold}CO$ monocarbonyls. These gem-dicarbonyls were also observed after dosing was stopped and the chamber pumped to UHV conditions, in agreement with the stability of dicarbonyl species inferred from the adsorption energy (Fig. S2). A blank experiment on clean $Fe_3O_4(001)$ showed no adsorption, confirming that the species observed originate from Rh.

Based on TPD results,[71] we estimate that the residence time of a CO molecule on the $Fe_3O_4(001)$ surface at RT is in the nanosecond regime. This implies that CO can diffuse in a weakly bound precursor state before being captured at a Rh site. This is consistent with the rapid saturation of the monocarbonyl peak observed in CO exposure-dependent IRAS spectra (66% saturation is achieved already at 0.063 L exposure, Figure S3). By fitting the uptake curve, we determined the effective capture area per $Rh_{2fold}$ site to be approximately 4 $nm^2$, which means that a large fraction of the CO arriving at the surface is efficiently captured at Rh sites in the initial regime. This is consistent with our STM observations, allowing for a moderate tip-shadowing effect (Figure 1 and Supplementary Movie S1). Since sequential CO adsorption to form $Rh_{2fold}(CO)_2$ species was not observed experimentally by STM under UHV conditions, even though CO encounters pre-existing $Rh_{2fold}CO$ roughly every 10 seconds at $10^{-8}$ mbar, we conclude that the capture of the second CO molecule on an isolated monocarbonyl is a rare event. This suggests that the second CO is not generally stabilised upon encounter with $Rh_{2fold}CO$ unless the local Rh–support geometry is transiently in a capture-ready configuration. Assuming the same 4 $nm^2$ capture zone, each $Rh_{2fold}CO$ experiences roughly $2\times10^7$ CO encounters per second at 2 mbar. Thus, the immediate observation of $Rh_{2fold}(CO)_2$ at 2 mbar CO in NAP-STM suggests that the increase in the encounter rate renders the requisite rare events accessible on the experimental timescale, leading to the formation of thermodynamically stable $Rh_{2fold}(CO)_2$ species.

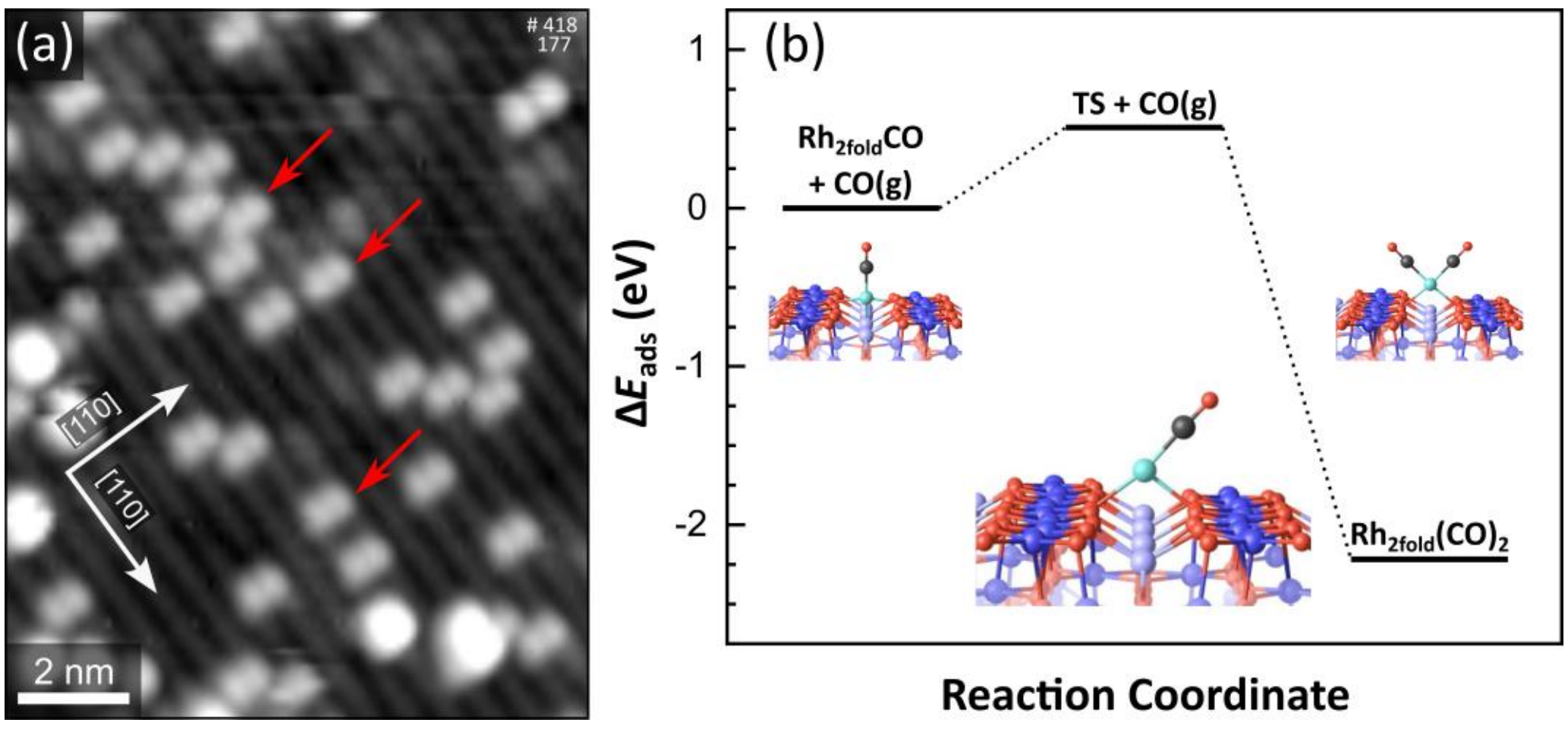

**Figure 5: Exposure of $Rh/Fe_3O_4(001)$ to 2 mbar CO.** (a) NAP-STM image ($V_{sample}$ = 2.2 V; $I$ = 0.02 nA) of 0.2 ML $Rh/Fe_3O_4(001)$ under 2 mbar CO. Two-lobe features are observed as the dominant species and are assigned to gem-dicarbonyls formed by adsorption of a second CO molecule on $Rh_{2fold}CO$ species (red arrows in Fig. 1a). (b) Adsorption of a second CO molecule onto the monocarbonyl configuration proceeds via a metastable transition state (TS), occupied ≈$10^{-8}$ or $10^{-9}$ of the time at room temperature, as deduced from DFT calculations in our previous study.[46] Energies are given relative to the monocarbonyl ground state.

## 5. Discussion

The results presented here demonstrate that IRAS experiments on single crystals under well-defined conditions can provide quantitatively reliable, site-resolved vibrational fingerprints for oxide-supported single-atom catalysts. Using Rh/$Fe_3O_4$(001) as an exemplary model system, we resolve three distinct Rh–CO species, namely monocarbonyls, and gem-dicarbonyls of twofold-coordinated Rh adatoms, as well as monocarbonyls formed at fivefold-coordinated Rh atoms embedded in the surface. We could assign the vibrational frequencies $\nu_{CO}$ unambiguously to these species by combining isotope substitution, annealing experiments, STM/nc-AFM imaging, XPS, TPD, and DFT-based calculations. The coordination-dependent shifts in $\nu_{CO}$ provide a direct handle on the local environment of Rh and can be used as a reference when analysing the more complex CO-IR spectra of supported Rh powder catalysts.[18,24,25,28,69,70]

The present study also exemplifies how single-crystal IRAS relates to CO-IR spectra on powders. On powder catalysts, randomly oriented crystallites and averaging over all incidence angles mean that both symmetric and asymmetric gem-dicarbonyl stretches typically appear as bands of increased absorption.[18,24,25,28,69,70] On the single-crystal Rh/$Fe_3O_4$(001) model system, by contrast, the fixed surface orientation and chosen incidence angle range make the IRAS intensity strongly dependent on dipole direction: components normal to the surface dominate, while in-plane components contribute only weak, oppositely signed features. This explains why we only observe the symmetric stretch of the gem-dicarbonyl and do not observe the asymmetric mode, even though both are prominent in powder spectra. While the spectra therefore differ in the intensities of the IR bands, the fingerprint positions of the IR-active species remain unaffected, making UHV-based IRAS studies valuable for identifying, assigning and benchmarking surface species.

The comparison with theory highlights both the strengths and limitations of current DFT approaches. DFT+U calculations correctly predict the relative ordering of $\nu_{CO}$ for the $Rh_{2fold}CO$ and $Rh_{2fold}(CO)_2$ species at twofold-coordinated Rh adatom sites, and the calculated adsorption energies rule out gem-dicarbonyl formation at $Rh_{5fold}$ sites. On the other hand, the calculated difference between the symmetric stretch of $Rh_{2fold}(CO)_2$ and the CO stretch frequency at $Rh_{5fold}$ strongly depends on details of the computational approach (whether a Hubbard *U* is used for the Rh), so assigning these bands purely on the basis of the calculated frequencies would be ambiguous. Since the CO frequency depends on the charge of the Rh,[66] we attribute this sensitivity to the different charge states of Rh at these sites: At twofold sites, $Rh_{2fold}$ is singly charged[46], while the calculated Bader charge of the $Rh_{5fold}$ is 1.27 *e* (1.23 *e* with $U_{eff,Rh}$ = 3.61 eV), indicating a $Rh^{2+}$ or $Rh^{3+}$ state. Usually, the $Rh^{3+}$ oxidation state is more common, but less likely to form stable carbonyls than $Rh^{2+}$.[66] In our calculations, the Hubbard *U* mainly shifts the $\nu_{CO}$ on the $Rh_{5fold}$, while the other configurations are much less affected. Employing *U* is common for similar systems with 3d transition elements,[47] but it is often neglected for the 4d and 5d metals since several works indicate that *U* values for 4d elements are lower and therefore of less importance.[72,73] Other work, however, indicates similar *U* values for Fe and Rh,[74] Given that, in the present system, using $U_{eff,Rh}$ = 3.61 eV significantly improves the agreement between calculated and experimental $\nu_{CO}$ at $Rh_{5fold}$, it may be worthwhile to re-examine the role of Hubbard corrections for 4d (and possibly 5d) elements more generally, even though a single case study is clearly insufficient to justify a universal prescription.

Since the configuration corresponding to each frequency can be unambiguously determined by our experiments, the experimental spectra provide a valuable benchmark for refining computational treatment of metal–CO bonding on oxides. More broadly, the combination of well-defined model surfaces, high-quality IRAS, and complementary microscopy across the UHV–near-ambient range offers a general strategy for understanding the CO-IR signatures of isolated metal sites in real catalysts. This allows identification of which structural motifs and formation pathways are actually relevant under working conditions.

A central mechanistic question concerns the formation of gem-dicarbonyls. Under UHV, gem-dicarbonyl $Rh_{2fold}(CO)_2$ species arise almost exclusively from CO-induced dissociation of $Rh_2$ dimers, even though the formation of a dicarbonyl from an isolated monocarbonyl is thermodynamically favourable. This behaviour reflects a strong kinetic constraint: sequential adsorption of a second CO molecule onto a $Rh_{2fold}CO$ appears to require that CO arrives when the Rh–CO complex is transiently in a capture-ready configuration, a circumstance that is rarely realised at the low encounter rates present during UHV dosing. In our previous work, [46] we identified one plausible example of such a capture-ready configuration in the form of a metastable geometry approximately 0.5 eV above the $Rh_{2fold}CO$ ground state [Fig. 5(b)]. An energy difference of 0.5 eV corresponds to a configurational probability on the order of $10^{-8}$–$10^{-9}$ at 300 K, which is consistent with the absence of sequential $Rh_{2fold}(CO)_2$ formation from isolated $Rh_{2fold}CO$ under UHV conditions, while at millibar CO pressures the dramatically increased encounter rate renders such rare configurations accessible on experimentally relevant timescales. As shown by the NAP-STM experiments in Section 5, this restriction is lifted at 2 mbar CO, where gem-dicarbonyls form directly on isolated $Rh_{2fold}$ sites and dominate the surface coverage. While the present data do not uniquely identify the capture-ready configuration and additional pathways may also contribute, the combination of UHV and NAP provides a coherent picture in which UHV acts as a kinetic filter that isolates individual reaction channels, whereas near-ambient measurements reveal which pathways dominate under conditions closer to applied catalysis.

## Conclusions

The results demonstrate how IRAS can serve as a bridging technique between idealised model systems and applied systems under realistic conditions. Using $Rh/Fe_3O_4(001)$ as a model system in UHV, we resolved and assigned the key Rh carbonyl species present on the surface: monocarbonyls and gem-dicarbonyls at twofold-coordinated Rh sites, as well as monocarbonyls formed by fivefold-coordinated Rh embedded in the surface. Under UHV conditions, gem-dicarbonyls are a minority species since they form almost exclusively via CO-induced dissociation of $Rh_2$ dimers, because the sequential adsorption of two CO molecules on an isolated $Rh_{2fold}$ is kinetically blocked despite being thermodynamically favoured. Near-ambient-pressure STM shows that at millibar CO pressures this kinetic limitation is lifted and gem-dicarbonyls form directly on isolated $Rh_{2fold}$ sites, reconciling UHV model studies with the gem-dicarbonyl signatures commonly observed under catalytic conditions. The experimental CO stretch frequencies are valuable as fingerprints for oxide-supported single-atom catalysts. Comparison with DFT+U highlights that theory does not yet match experimental frequencies quantitatively, underlining the value of the IRAS benchmarks established here for interpreting CO-IR spectra and refining theoretical descriptions of metal–CO bonding on oxide-supported single-atom catalysts.

## Data availability

The data supporting this article have been included as part of the Supplementary Information.

## Author contributions

NEHB contributed to the investigation, performed formal analysis, data curation, visualisation, and prepared the original draft. CW assisted with data analysis and investigation. PS carried out the theoretical and methodological work, contributed to formal analysis and visualisation. DR supported methodology and contributed to experimental investigation. AL, FL, MO, and LP performed experimental investigation. ZJ contributed to investigation and formal analysis. FK performed data analysis and visualisation. ME contributed to methodology. MS, UD, MM, and CF contributed to the methodology and validation. PM contributed to investigation and methodology. JP contributed to conceptualisation, supervision, validation, and funding acquisition. GSP contributed to conceptualisation, supervision, validation, funding acquisition, and writing of the original draft. All authors participated in reviewing & editing of the manuscript.

## Conflicts of interest

There are no conflicts to declare.

## Acknowledgements

Funding from the European Research Council (ERC) under the European Union's Horizon 2020 research and innovation programme (grant agreement No. [864628], Consolidator Research Grant 'E-SAC') is acknowledged. This research was funded in part by the Austrian Science Fund (FWF) 10.55776/F8100 and the Cluster of Excellence MECS (10.55776/COE5). Jiri Pavelec and Adam Lagin gratefully acknowledge financial support from the FWF through project number 10.55776/I6732. Chunlei Wang gratefully acknowledges financial support from the FWF through project number 10.55776/PAT1934924. Matthias Meier gratefully acknowledges financial support from the FWF through project number 10.55776/PAT2176923. Moritz Eder acknowledges funding by the Marie Skłodowska-Curie Actions (Project 101103731, SCI-PHI). The authors acknowledge the CERIC-ERIC consortium (proposal ID 20252185) for access to experimental facilities and financial support. Peter Matvija acknowledges the financial support from the Ministry of Education, Youth, and Sports of the Czech Republic under project LM2023072. The Austrian Scientific Computing was used to obtain the computational results. 

## Notes and references

a. Institute of Applied Physics, TU Wien, 1040 Vienna, Austria

b. Department of Surface and Plasma Science, Charles University in Prague, V Holesovickach 2, 180 00 Praha, Czech Republic

c. Central European Institute of Technology (CEITEC), Brno University of Technology, Brno, 612 00, Czechia

d. Faculty of Physics and Center for Computational Materials Science, University of Vienna, Vienna AT 1090, Austria

e. Dipartimento di Fisica e Astronomia, Università di Bologna, Bologna IT 40126, Italy

* jiri.pavelec@tuwien.ac.at

## Supporting Information

**Figure S1. Normalised reflectivity difference $\Delta R/R_0$ for an adsorbate on $Fe_3O_4$ at a representative**

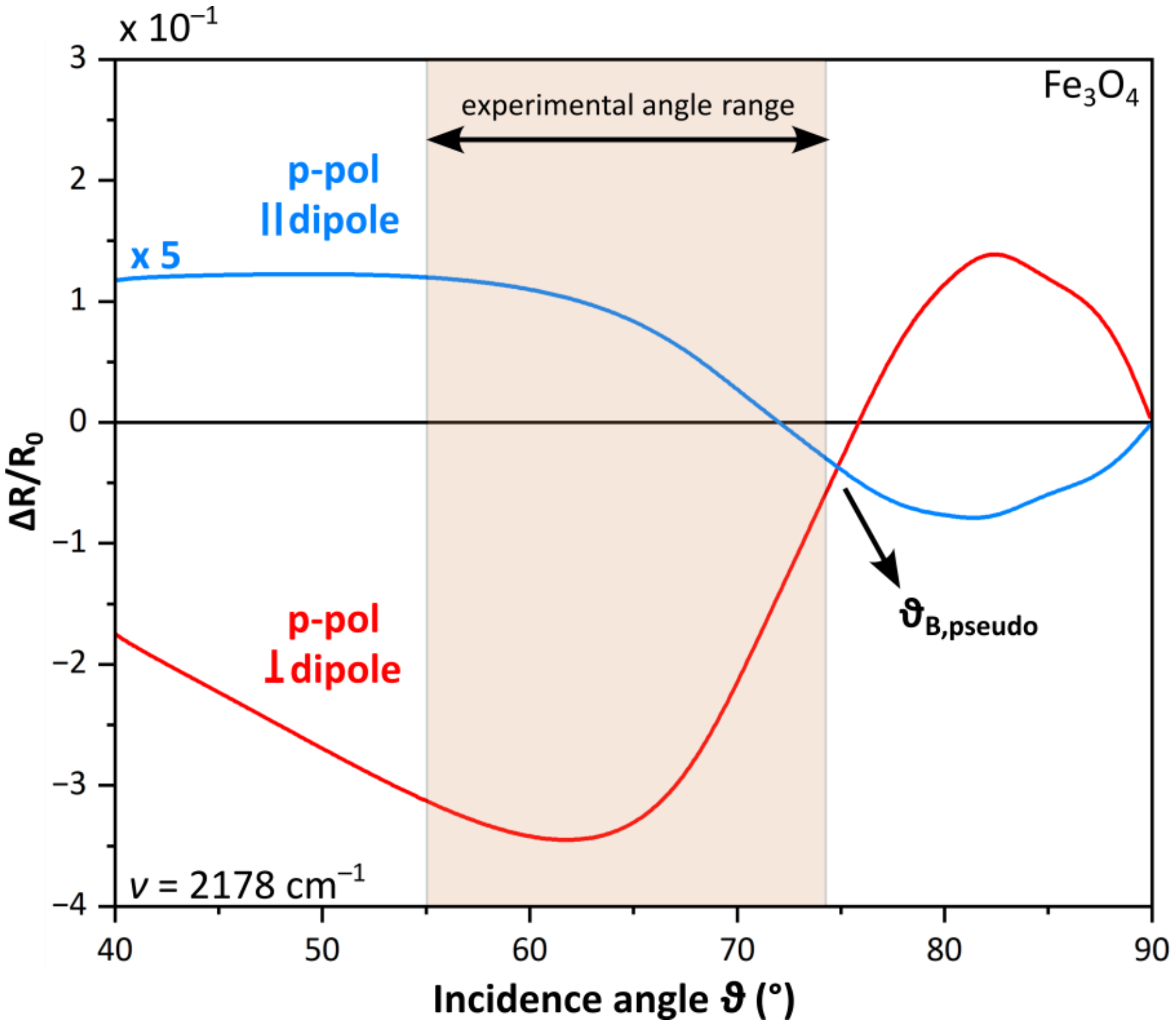

**frequency of $v$ = 2178 cm⁻¹, including system-specific corrections.** $\Delta R/R_0$ was calculated using a static electronic polarisability $\alpha_e = 0$, and a vibrational polarisability $\alpha_v = 3.4 \times 10^{-26}$ cm$^3$. The calculation includes corrections for an incidence-angle spread of $\kappa = \pm 2°$, a system-specific depolarisation factor $\Gamma = 0.21$, and the contribution from vicinity illumination.[1,2] The red curve shows the response for p-polarised light interacting with a dipole moment perpendicular to the surface, while the blue curve corresponds to p-polarised light with a dipole moment parallel to the surface and lying in the plane of incidence (scaled by a factor of 5 for clarity). The beige region indicates the incidence-angle range used in our experiments (55°–74°), which lies below the pseudo-Brewster angle where the characteristic band inversion occurs. Within this range, perpendicular dipoles give rise to strong negative peaks, whereas parallel dipoles yield much weaker positive features—approximately fifteen times smaller in magnitude. The signal in p-polarisation vanishes for a dipole moment normal to the plane of incidence. Since the azimuthal orientation of alternating (001) planes of $Fe_3O_4$(001) changes by 90°, only half of the molecules with a given azimuthal orientation with respect to the Fe rows contribute to the signal. This leads to an effective factor of 30 in peak height between out-of-plane and in-plane dipole moments.

**Table S1.** Relative increase in IRAS peak intensities between 0.2 ML and 0.4 ML Rh/$Fe_3O_4$(001)

| Peak Position | Peak height (0.2 ML Rh) | Peak height (0.4 ML Rh) | Relative increase |
|---|---|---|---|
| 1979 | $-5.4 \times 10^{-4}$ | $-7.9 \times 10^{-4}$ | ≈1.5 |
| 2037 | $-8.8 \times 10^{-5}$ | $-3.8 \times 10^{-4}$ | ≈4.3 |
| 2059 | $-5.9 \times 10^{-5}$ | $-3.0 \times 10^{-4}$ | ≈5.1 |

The table summarises the change in IRAS peak intensities between Rh coverages of 0.2 ML (black spectrum in Fig.2) and 0.4 ML (blue spectrum in Fig. 2). The band at 1979 cm$^{-1}$, associated with CO bound to twofold coordinated $Rh_{2fold}$ sites, increases by a factor of ≈1.5. In comparison, the gem-dicarbonyl symmetric stretch at 2037 cm$^{-1}$ increases by a factor of ≈4.3, and the band at 2059 cm$^{-1}$ attributed to CO adsorption on $Rh_{5fold}$ sites increases by a factor of ≈5.1. The substantially larger relative increases of the higher-frequency bands further support their assignment to gem-dicarbonyl species and CO bound to $Rh_{5fold}$ sites.

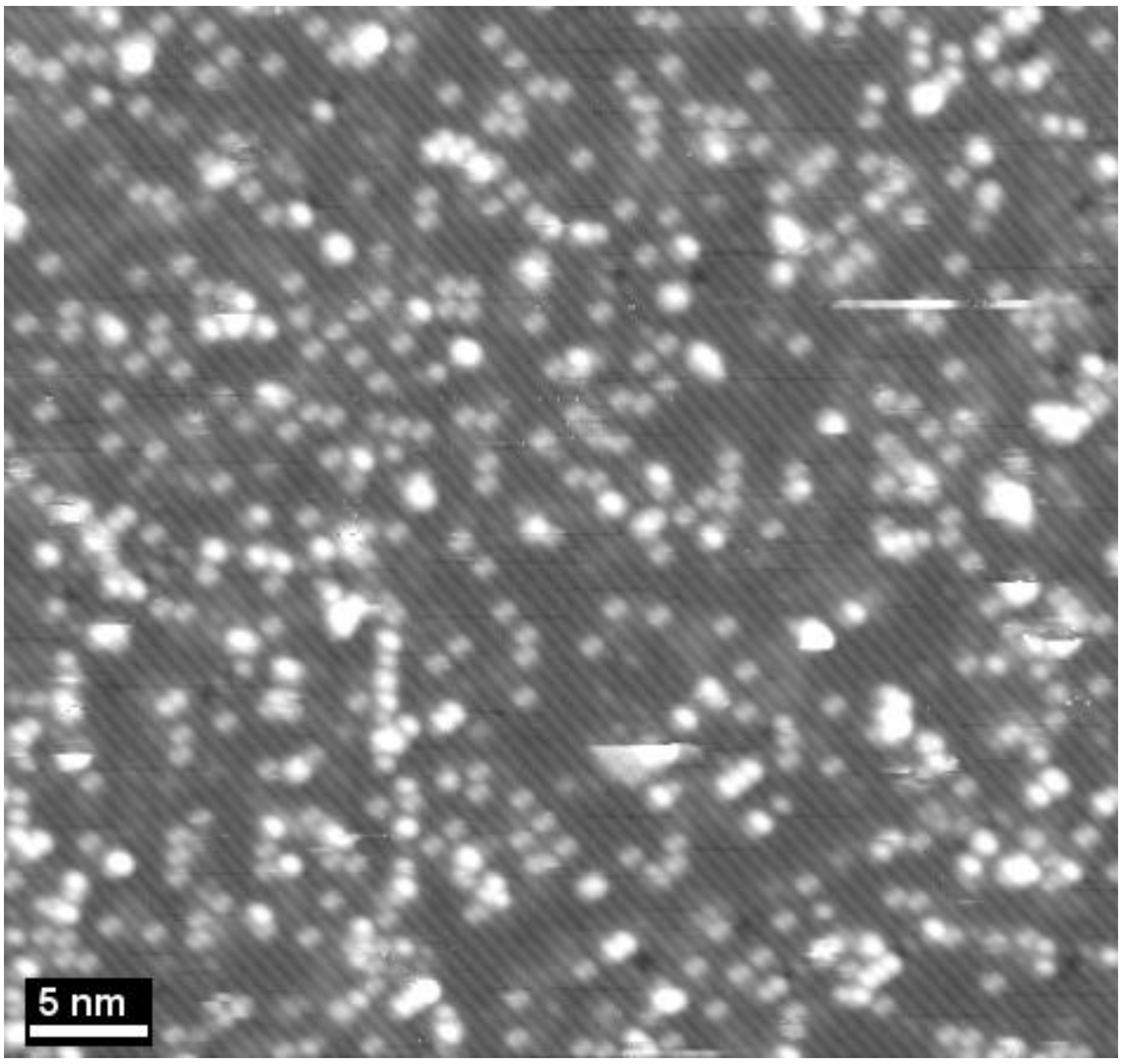

**Figure S2. STM measurement of CO adsorption on Rh/$Fe_3O_4$(001).** This STM image was acquired under UHV conditions following the NAP-STM measurement of Rh/$Fe_3O_4$(001) in 2 mbar CO shown in Figure 5a. The dicarbonyl species formed during CO exposure remain stable even under UHV, consistent with the thermodynamic stability predicted by DFT (adsorption energy: −2.21 eV per CO). The few clusters or blurry features visible in the image may arise from minor contamination deposited during the prolonged CO exposure at 2 mbar.

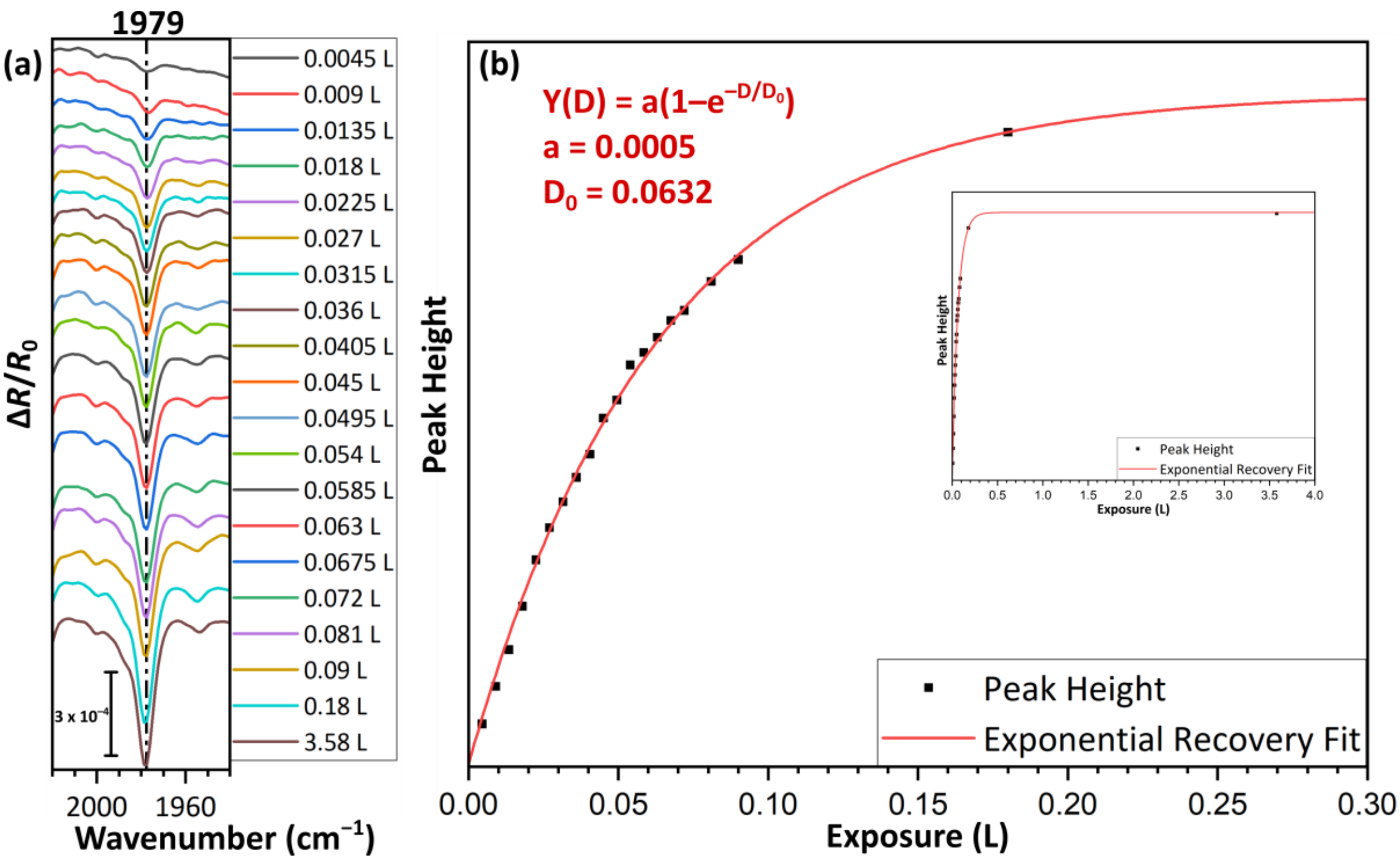

**Figure S3. CO exposure-dependence of the $Rh_{2fold}CO$ IRAS band.** (a) p-polarised IRAS spectra of a Rh-decorated $Fe_3O_4$(001) surface with a Rh coverage of 0.2 monolayer (ML; 1 ML = 1.42 × $10^{14}$ Rh atoms per $cm^2$)

upon exposure to increasing doses of $^{13}CO$. The reference spectrum $R_0$ was recorded after Rh deposition and prior to the first CO dose of 0.0045 Langmuir (L; 1 L = 1.33 x $10^{-6}$ mbar.s) and was used as a background for all spectra. Already after the first exposure, a weak $Rh_{2fold}CO$ band is visible. With successive 0.0045 L increments, the $Rh_{2fold}CO$ peak increases progressively in intensity and approaches saturation at a total exposure of ≈0.18 L. (b) $Rh_{2fold}CO$ IRAS peak height as a function of CO exposure (black squares), together with an exponential recovery fit (red curve) of the form $y(D) = a[1 - exp(-D/D_0)]$ from which a value of $D_0$ = 0.0632 is extracted and used to estimate the effective capture area of $Rh_{2fold}$ sites.

The adsorption of CO on Rh/$Fe_3O_4$(001) was quantified by monitoring the IRAS band intensity of $Rh_{2fold}CO$ as a function of CO exposure $D$ (in Pa.s). The approach to saturation is well described by a single-exponential function, $y(D) = a[1 - exp(-D/D_0)]$, where $D_0$ is the characteristic exposure at which $1 - 1/e$ (63%) of the total maximum peak height is reached.
This behaviour is consistent with first-order adsorption within a Langmuir-type model, for which the fractional coverage of $Rh_{2fold}CO$ sites is given by

$$\theta(D) = 1 - \exp\left[-S_0 I D/(P n_a)\right],$$

where $\theta$ is the fraction of $Rh_{2fold}$ sites occupied by CO, $S_0$ is the initial sticking coefficient, $I$ is the CO impingement rate (number of CO molecules per unit area and time), and $n_a$ is the number of Rh atoms per area. We express the initial sticking coefficient with an effective capture area $A_c$ per Rh atom as $S_0 = A_c n_a$. This effective capture area can be much larger than the geometric size of a Rh atom if the CO molecule can diffuse as a precursor before being captured by a Rh atom. Comparison with the exponential fit yields $A_c = P/(D_0 I)$. This comparison is valid because the $Rh_{2fold}CO$ species are well isolated under the present conditions, and the IRAS peak intensity is proportional to the CO coverage on $Rh_{2fold}$ sites.
From kinetic gas theory, the CO impingement rate is given by

$$I = P/\sqrt{2\pi m k_B T},$$

where $m$ is the mass of a CO molecule, $k_B$ is the Boltzmann constant, and $T = 300$ K. This gives $A_c = \sqrt{2\pi m k_B T}/D_0$. Using the experimentally determined value $D_0 = 0.063\ L = 8.38 \times 10^{-6}$ Pa.s, we obtain an effective capture area of $A_c = 4.1$ nm$^2$.
This analysis indicates that CO adsorption on $Rh_{2fold}$ sites is highly efficient, with each Rh atom effectively capturing CO molecules incident within its associated capture area.

When compared to the geometric area per Rh atom estimated from the nominal Rh density at 0.2 ML coverage ($n_{\mathrm{Rh}} = 0.2$ ML $\times 1.42 \times 10^{14}$ cm$^{-2}$), which corresponds to an average area of 3.52 nm$^2$ per Rh atom, the extracted capture area is slightly larger. This would correspond to a sticking coefficient of 1.16, which is unphysical. We attribute this discrepancy mainly to the uncertainty of the Rh coverage deposited (10–20%) and of the CO dose (≈10%). In addition, not all Rh atoms are isolated adatoms in twofold sites, but only the $Rh_{2fold}$ contributes to the IRAS peak at 1979 cm$^{-1}$. In any case, the Langmuir model is inaccurate for precursor-mediated adsorption with a capture area close to the area per adsorption site. Nevertheless, since the fit of the IRAS intensities vs. dose assuming Langmuirian adsorption is reasonable, the capture area derived from this model should be useful as a rough estimate. If the effective capture area were much larger, i.e. if the precursor molecules could visit many Rh sites before leaving the surface, the IRAS peak height as a function of dose would not follow the exponential recovery function observed in experiment but increase almost linearly up to saturation.

In principle, one could also try to estimate the CO sticking probability from the STM movies. This would reduce the uncertainty in the density of adsorption sites $n_a$, since one could count these sites in the STM images. Nevertheless, this method of estimating the sticking probability is very unreliable, since shadowing by the tip strongly reduces the local impingement rate in the area scanned.